\documentclass[preprint,12pt]{elsarticle}

\usepackage{amssymb}
\usepackage{graphicx}
\usepackage{epsfig}
\usepackage{graphics}
\usepackage{amsmath}

\newcommand{\Prime}{''}

\newcommand{\ux}{\hat{x}}
\newcommand{\uy}{\hat{y}}
\newcommand{\uz}{\hat{z}}

\newcommand{\woo}{\omega_{0}^{\prime}}

\newcommand{\wrf}{\omega_{rf}}

\begin{document}

\begin{frontmatter}



\title{Simultaneous $\pi/2$ rotation of two spin species of different 
gyromagnetic ratios}

\author[duke,uiuc]{Ping-Han~Chu\corref{mycorrespondingauthor}}
\cortext[mycorrespondingauthor]{Present address: Los Alamos National Lab, P.O. Box 1663, H803, Los Alamos, NM, 87544, USA. Tel.: +1 5056060510}
\ead{pchu@lanl.gov}
\author[uiuc]{Jen-Chieh~Peng}

\address[duke]{Triangle Universities Nuclear Laboratory and Department of Physics, Duke University, Durham, North Carolina 27708, USA}
\address[uiuc]{Department of Physics, University of Illinois at Urbana-Champaign, Urbana, Illinois 61801, USA}

\begin{abstract}
We examine the characteristics of the $\pi/2$ pulse for simultaneously rotating two spin species of different gyromagnetic ratios with the same sign. For a $\pi/2$ pulse using a rotating magnetic field, we derive the equation relating the frequency and strength of the pulse to the gyromagnetic ratios of the two particles and the strength of the constant holding field. For a $\pi/2$ pulse using a linear oscillatory magnetic field, we obtain the solutions numerically, and compare them with the solutions for the rotating $\pi/2$ pulse. Application of this analysis to the specific case of rotating neutrons and $^3$He atoms simultaneously with a $\pi/2$ pulse, proposed for a neutron electric dipole moment experiment, is also presented.
\end{abstract}


\end{frontmatter}



The $\pi/2$ rotation is a commonly used technique in nuclear magnetic resonance. A spin $\vec S$ with a gyromagnetic ratio $\gamma$, pointing along a constant magnetic field $B_0$ in the $\hat z$ direction, can be rotated into the $\hat x - \hat y$ plane by applying a radio-frequency (rf) magnetic field $\vec{B}(t)$, called a $\pi/2$ pulse, in the direction orthogonal to $B_0$. For a $\pi/2$ pulse rotating in the $\hat x - \hat y$ plane, $\vec{B}(t)$ can be written as
\begin{align}
B_x(t)&=B_1\sin{(\wrf t)},~B_y(t)=B_1\cos{(\wrf t)},
\label{eq:b1}
\end{align}
\noindent which represents a magnetic field with frequency $\wrf$ and an amplitude $B_1$. While $\omega_{rf}$ is often chosen to be the same as the resonant Larmor frequency, $\omega_0 = \gamma B_0$,  off-resonance frequencies could also be used. As discussed below, this capability to perform a $\pi/2$ rotation with off-resonance frequencies makes it possible to rotate simultaneously two spin species of different gyromagnetic ratios with a single $\pi/2$ pulse of suitable frequency and duration.

The need to rotate two different spin species with a single $\pi/2$ pulse 
is relevant for the proposed experiment~\cite{Golub:1994cg,nEDM:2002,Ito:2007} 
at the Oak Ridge National Laboratory to search for the neutron electric 
dipole moment (EDM) using polarized ultra-cold neutrons stored in a 
superfluid helium cell containing polarized $^3$He acting as 
a co-magnetometer and a spin-analyzer. A non-zero neutron EDM can be 
observed by measuring the difference of the precession frequencies 
of neutrons when electric and magnetic fields are aligned or anti-aligned. 
The initial polarization directions for neutron and $^3$He spins are 
parallel to a constant $B_0$ field pointing along the $\hat z$ axis. 
To measure the precession frequency of neutrons relative to that of $^3$He, 
a single $\pi/2$ pulse will be applied to rotate both spins into the 
$\hat x - \hat y$ plane.
A superconducting quantum interference devices (SQUID) will be used to
measure the precession frequency of $^3$He~\cite{Kim:2013}, and the 
relative precession between neutrons and $^3$He can be determined by the 
rate of the absorption reaction, n + $^3$He $\rightarrow$ p + $^3$H, 
which depends sensitively on the relative spin orientations of neutrons
and $^3$He~\cite{Passel:1966}. From the measurements of the precession
of $^3$He and the absorption reaction of n-$^3$He, the precession
frequency of neutrons can be determined.
An alternative method, the dressed spin technique, could also be 
used to measure the neutron EDM by applying an additional
dressing field~\cite{Golub:1994cg,Chu:2011}.

It is not evident that two spin species with different gyromagnetic
ratios could be rotated simultaneously by a $\pi$/2 pulse. 
As shown in this paper, this is only possible if the ratio of the
magnetic moments of the two spin species falls within a certain
range. Vasserman et al. utilized a method to simultaneously rotate 
the spins of electron and positron by $90^\circ$ to measure their 
anomalous magnetic moments for a test of CPT 
invariance~\cite{Vasserman:1987}. This method is only 
applicable for particles with identical magnitude of gyromagnetic ratios.
Recently, de Lange et al.
published a new method to manipulate two spin species using the 
spin echo technique~\cite{Lange:2012}.
However, this method requires a spin bath which does not exist 
in many experiments such as the neutron EDM experiment. It is necessary
to consider other methods which do not require a spin bath.
The purpose of this
paper is to discuss a method which is generally applicable to 
experiments requiring a simultaneous $\pi$/2 rotation for two 
different spin species. 

The rest of this paper is organized as follows.
First, we derive the equations for calculating $\omega_{rf}$ and $\tau$ as a function of $B_0$, $B_1$, and the gyromagnetic ratios of two different 
spin species. Second, we explore the range of these parameters and the 
relation between them. Third, we present the numerical solutions and comment 
on the characteristics of the solutions. Finally, we also consider the case 
for linear rf magnetic fields, which are more readily implemented than 
rotating fields, and compare the solutions with those of rotating rf fields. 
The application of this study to the specific case of the  neutron 
EDM experiment involving neutron and $^3$He will also be presented.
Some initial results obtained by one of the coauthors (JCP) for the case 
of rotating rf fields were presented in an unpublished report~\cite{nEDM:2002}. 
\begin{figure}[t]
\centering
\includegraphics[width=0.9\textwidth,height=0.4\textheight]{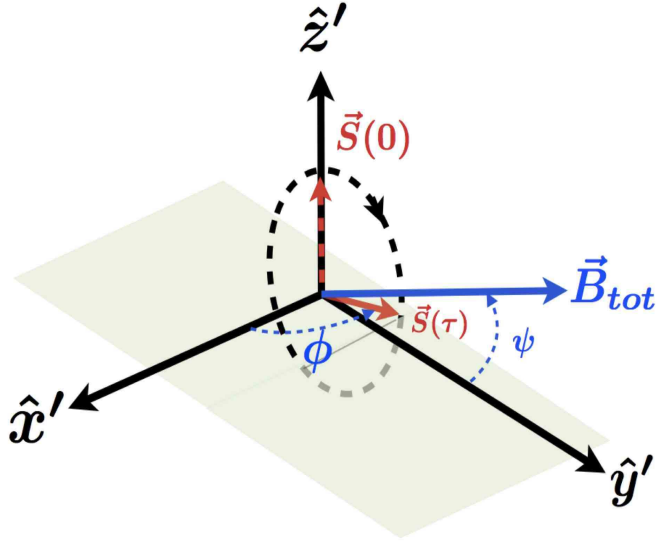}
\caption{ (Color figure) Schematic plot of spin rotation using a rotating rf field $B_1(t)$ in the $\ux-\uy$ plane with frequency $\omega_{rf}$. The holding field $B_0$ is along $\uz$-axis. In a frame rotating  at $\omega_{rf}$ along $\uz$, the effective field is $\vec{B}_{tot} = (B_0-\frac{\omega_{rf}}{\gamma})\hat{z}^{\prime} + B_1\hat{y}^{\prime}$. $\vec{S}(0)$ is the initial spin orientation and $\vec{S}(\tau)$ is the spin orientation after applying a $\pi/2$ pulse. $\psi$ is the angle between $\vec{B}_{tot}$ and $\hat{y}^{\prime}$ and $\phi$ the angle between $\vec{S}(\tau)$ and $\hat{x}^{\prime}$.}
\label{fig:schematic}
\end{figure}

\begin{figure}[h]
\centering
\includegraphics[width=0.8\textwidth, height=0.4\textheight]{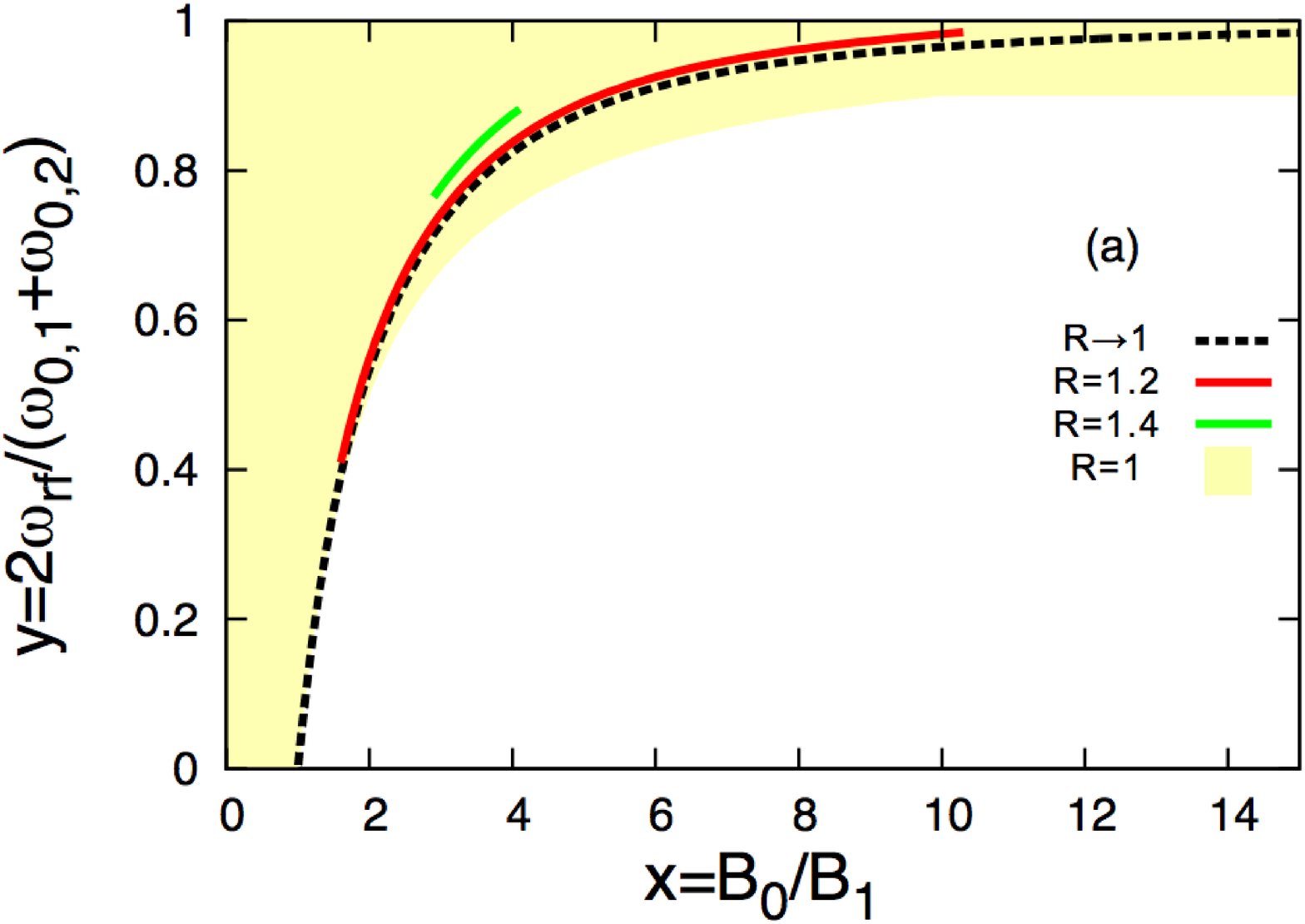}
\includegraphics[width=0.8\textwidth, height=0.4\textheight]{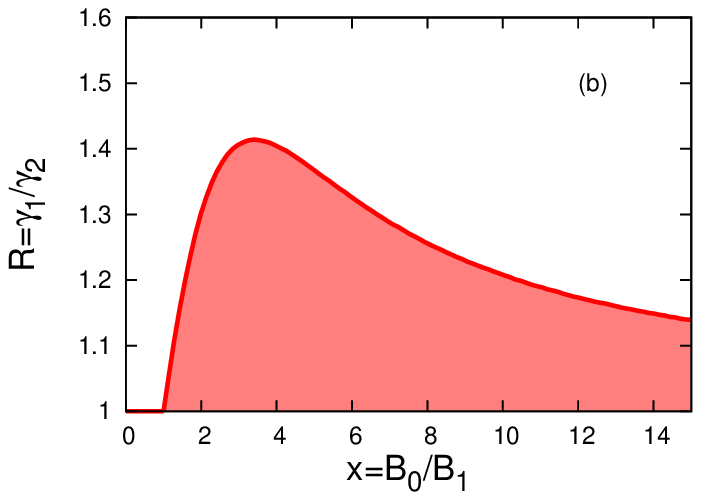}
\caption{(Color figure) Solutions of the $\pi/2$ rotation for two spin species. (a) $y=2\wrf /(\omega_{0,1}+\omega_{0,2})$ versus $x=B_0/B_1$ for different $R = \gamma_1/\gamma_2$. The yellow region shows the constraint of the $\pi/2$ rotation for a single spin species according to Eq.~\ref{eq:constraint}. (b) $R=\gamma_1/\gamma_2$ versus $x=B_0/B_1$. All values in the red region can be used for $\pi/2$ pulses. }
\label{fig:cross.xy.n}
\end{figure}

In general, the spin motion of polarized particles moving in an external electromagnetic field can be described by the Bargmann-Michel-Telegdi equation~\cite{Bargmann:1959}. However, for non-relativistic particles, the dynamics of a spin 
in a magnetic field system can be described by the Bloch equation~\cite{Bloch:1946}:
\begin{align}
\frac{d\vec{S}}{d t} &= \vec{S} \times (\gamma \vec{B}(t)).
\label{eq:bloch}
\end{align}
Here we consider a system of polarized particles moving slowly in a cell with a static magnetic field, $B_0$, parallel to the spin orientation of polarized particles along the $\uz$-axis. Upon the application of the rotating field given in Eq.~\ref{eq:b1}, the effect on the spin direction of the particles can be conveniently described by transforming to a frame rotating at $\wrf$ along $\uz$. In this rotating frame with axes defined by the $\ux^\prime, \uy^\prime,$ and $\uz^\prime$ as shown in Fig.~\ref{fig:schematic}, the field $\vec B(t)$ becomes static with a magnitude $B_1$ pointing along the $\hat y^{\prime}$ axis. The static $B_0$ field in the $\hat z^{\prime}$ axis is changed to $B_0 - \wrf/\gamma$ in this rotating frame and the spins of the particles will precess about the magnetic field $\vec B_{tot}$ given as
\begin{equation}
\vec{B}_{tot}=(B_0-\frac{\wrf}{\gamma}) \hat z^{\prime} 
+ B_1 \hat y^{\prime}
\end{equation}
\noindent with an effective Larmor frequency 
\begin{align}
\omega^\prime_0 &= \gamma \sqrt{(B_0-\frac{\wrf}{\gamma})^2+B_1^2}.
\label{eq:omegaprime}
\end{align}

The angle $\psi$ between the directions of  $\vec{B}_{tot}$ and $\uy ^\prime$ satisfies
\begin{align}
\tan\psi = \frac{B_0-\wrf/\gamma}{B_1} = \frac{B_0}{B_1} 
(1-\frac{\wrf}{\gamma B_0}).
\label{eq:tanpsi}
\end{align}
\noindent It is convenient to define another coordinate system ($\ux^{\Prime}$, $\uy^{\Prime}$, $\uz^{\Prime}$) obtained from a rotation along the $\ux^{\prime}$ axis by an angle $\psi$ such that $\uy^{\Prime}$ is along the direction of $\vec B_{tot}$:
\begin{align}
\left[
\begin{array}{c}
\ux^{\Prime} \\
\uy^{\Prime} \\
\uz^{\Prime}
\end{array}
\right]&=\left[
\begin{array}{ccc}
1 & 0 & 0 \\
0 & \cos\psi & \sin\psi \\
0 & -\sin\psi & \cos\psi
\end{array}
\right]
\left[
\begin{array}{c}
\ux^{\prime} \\
\uy^{\prime} \\
\uz^{\prime}
\end{array}
\right].
\label{eq:5}
\end{align}
The spin of the particle is initially along the $\uz^\prime$-axis. At a later time $t$, it can be expressed as
\begin{equation}
\vec{S}(t) = \vec{S}_{\parallel}(t)+\vec{S}_{\bot}(t),
\end{equation}
where $\vec{S}_{\parallel}(t)$ is the component parallel to
$\vec{B}_{tot}$ and is independent of time,
\begin{align}
\vec{S}_{\parallel}(t) &=| \vec{S}|\sin\psi\uy^{\Prime}.
\end{align}
$\vec{S}_{\bot}$ is the component perpendicular to $\vec B_{tot}$ and will rotate about $\vec{B}_{tot}$ so that
\begin{equation}
\vec{S}_{\bot}(t) = (| \vec{S}| \cos\psi)(\cos{(\woo t)}\uz^{\Prime}
-\sin{(\woo t)}\ux^{\Prime}).
\label{eq:8}
\end{equation}
Using Eqs.~\ref{eq:5}-~\ref{eq:8}, $\vec{S}(t)$ can be expressed in terms of the coordinates in the rotating frame as 
\begin{align}
\vec{S}(t) & = \vec{S}_{\parallel}(t)+\vec{S}_{\bot}(t) \notag\\
&= |\vec{S}|(\sin\psi)\uy^{\Prime} + (|\vec{S}| 
\cos\psi)(\cos{(\woo t)}\uz^{\Prime}-\sin{(\woo t)}\ux^{\Prime}) \notag\\
&=|\vec{S}|[ -\cos\psi \sin({\woo t}) \ux^\prime+
\sin\psi \cos\psi(1 - \cos{\woo t})\uy^\prime
 \notag\\
&+(\sin^2\psi+\cos^2\psi\cos\woo t )\uz^\prime].
\label{eq:rfrotation}
\end{align}
For an rf pulse to rotate the spin from the vertical axis to the horizontal plane after a duration $\tau$, the component of $\vec{S}(\tau)$ along $\uz^\prime$ must vanish:
\begin{align}
&\vec{S}(\tau)\cdot \uz^\prime = \sin^2\psi+\cos^2\psi\cos\woo \tau = 0.
\label{eq:coswt}
\end{align}

We can readily obtain the solutions for $\omega_{rf}$ and $\tau$ to rotate two spin species of different gyromagnetic ratios by $\pi/2$ simultaneously. Due to the property of the $\cos$ function, the sign of $\omega_{0}^\prime$ does not change the solution of $\tau$; this implies the solution of $\tau$ is suitable for both positive and negative $\omega_{0}^\prime$ as well as $\gamma$. Denoting the gyromagnetic ratios for the two spin species as $\gamma_1$ and $\gamma_2$, and their Larmor frequencies as $\omega_{0,1}=\gamma_1 B_0$ and $\omega_{0,2}=\gamma_2 B_0$, the requirement is to have the same duration, $\tau$, with a given frequency to simultaneously rotate both species by $\pi/2$. Therefore, we obtain:
\begin{align}
\tau = \frac{\cos^{-1}(-\tan^2(\psi_1))}{|\omega^\prime_{0,1}|}
=\frac{\cos^{-1}(-\tan^2(\psi_2))}{|\omega^\prime_{0,2}|}
\label{eq:tau1}
\end{align}
where $\omega^{\prime}_{0,i}$ and $\psi_i$ are the effective Larmor frequency and $\psi$ angle for species $i$ in Eq.~\ref{eq:omegaprime} and Eq.~\ref{eq:tanpsi}, respectively. The range of $\cos^{-1}(-\tan^2(\psi))$ is from 0 to $\pi$ in order to keep $\tau$ positive.

The solutions for $\omega_{rf}$ must also satisfy the following constraint:
\begin{align}
&-1 \le \tan{\psi_i} \le 1,\notag
\end{align}
i.e.
\begin{align}
&-1 \le (\frac{B_0}{B_1})(1-\frac{\omega_{rf}}{\gamma_i B_0}) \le 1,
\end{align}
implying
\begin{align}
1 -  \frac{B_1}{B_0} \le \frac{\omega_{rf}}{\gamma_i B_0} \le 1 + \frac{B_1}{B_0}.
\label{eq:constraint}
\end{align}
 
It is also noted that after the application of a $\pi/2$ pulse, the spin directions of the two species on the $\hat x - \hat y$ plane would in general be different. Using Eqs.~\ref{eq:rfrotation} and~\ref{eq:coswt}, the azimuthal angle $\phi_i$ of the spin for species $i$ after a $\pi/2$ rotation is given as
\begin{align}
\phi_i &= 
\tan^{-1}(\frac{\sin\psi_i\cos\psi_i(1-\cos\omega^{\prime}_{0,i}\tau)}
{-\cos\psi_i\sin\omega^{\prime}_{0,i}\tau})
= -\tan^{-1}(\frac{1}{\sqrt{\cot^2\psi_i - 1}}).
\label{eq:phi}
\end{align}
Therefore, the spins of the two species are usually not aligned ($\phi_1 \ne \phi_2$) right after the $\pi/2$ pulse.
 
Equation~\ref{eq:tau1} can be solved numerically to find $\omega_{rf}$ 
for given values of $B_0, B_1,$ and $\gamma_i$. Because the gyromagnetic 
ratios of neutron and $^3$He are both negative, we focus on the case of 
both spin species having gyromagnetic ratios of the same sign in this 
paper. We note that 
Eq.~\ref{eq:tau1} is also applicable for the case of two spin species 
having gyromagnetic ratios of opposite signs. For convenience, we define dimensionless parameters if both gyromagnetic ratios are in the same sign: $R=\frac{\gamma_1}{\gamma_2}$, $x=\frac{B_0}{B_1}$ and $y =\frac{2\wrf}{\omega_{0,1}+\omega_{0,2}}$. Fig.~\ref{fig:cross.xy.n}(a) shows the solutions for $y$ as a function of $x$ for three different values of $R$. We only consider the case for $R \ge 1$, since the $R \le 1$ case simply corresponds to interchanging the two spin species. We first note that for a given value of $R~(R\equiv \gamma_1/\gamma_2)$, a solution for $\omega_{rf}$ can be found only within a certain range of $x$. This range becomes narrower as $R$ increases, and above certain value of $R$, there is no longer a solution for $\omega_{rf}$. The domain in $R$ versus $x$ for which a solution for $\omega_{rf}$ exists is shown as the red region in Fig.~\ref{fig:cross.xy.n}(b). 

Figure~\ref{fig:cross.xy.n}(a) also shows that the solutions for $y$ at different values of $R$ have very similar shapes. It is interesting to study the phenomena when $R$ approaches 1 (but not equal to 1). Inserting $R\approx 1 + \delta R$ into Eq.~\ref{eq:tau1} and ignoring higher-order terms in $\delta R$,
the solution for $y$ versus $x$ when $R\rightarrow 1$ is 
\begin{align}
&\cos^{-1}(x^2(-1+y)^2) \notag\\
=& \pi - \frac{2(x^2 y(1 - y))}{1+x^2(1-y)}\sqrt{\frac{1+x^2(1-y)^2}{1-x^2(1-y)^2}},
\label{eq:r=1}
\end{align}
shown as the black-dashed curve in Fig.~\ref{fig:cross.xy.n}(a). However, when $R=1$, corresponding to the degenerate case of a $\pi/2$ rotation for only a single species, the solution is given by Eq.~\ref{eq:constraint}, shown as the yellow region in Fig.~\ref{fig:cross.xy.n}(a). This abrupt change from a curve to a band for the solution of Eq.~\ref{eq:tau1} reflects the degeneracy occurring at $R = 1$.

\begin{figure}[h]
\centering
 \includegraphics[width=0.48\textwidth, height=0.3\textheight]{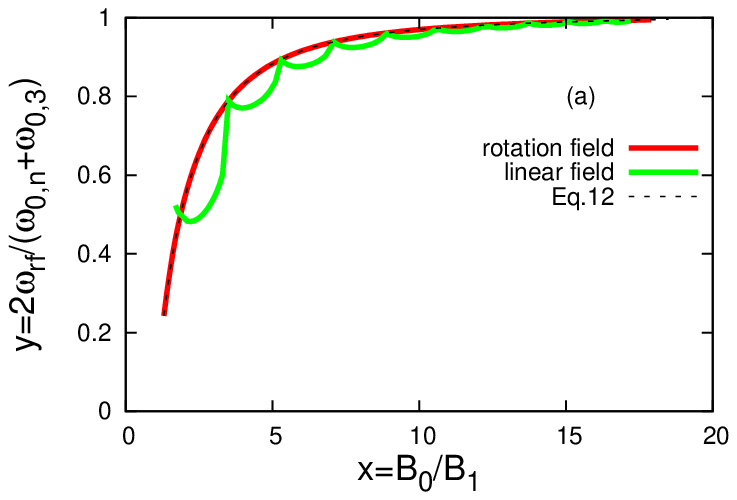}
 \includegraphics[width=0.48\textwidth, height=0.3\textheight]{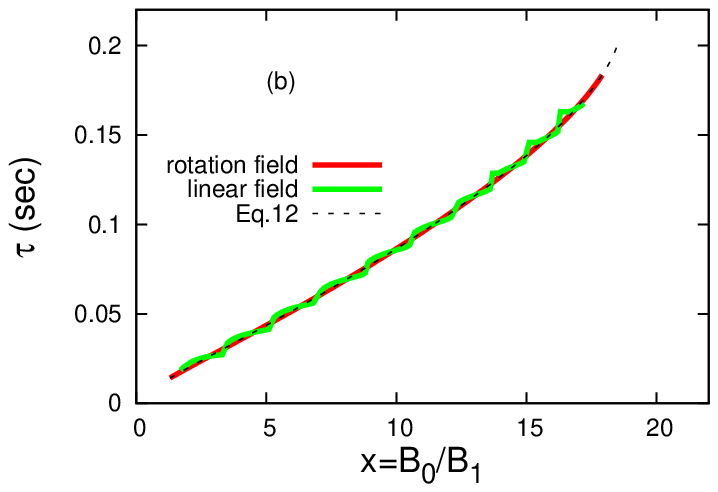}
 \includegraphics[width=0.48\textwidth, height=0.3\textheight]{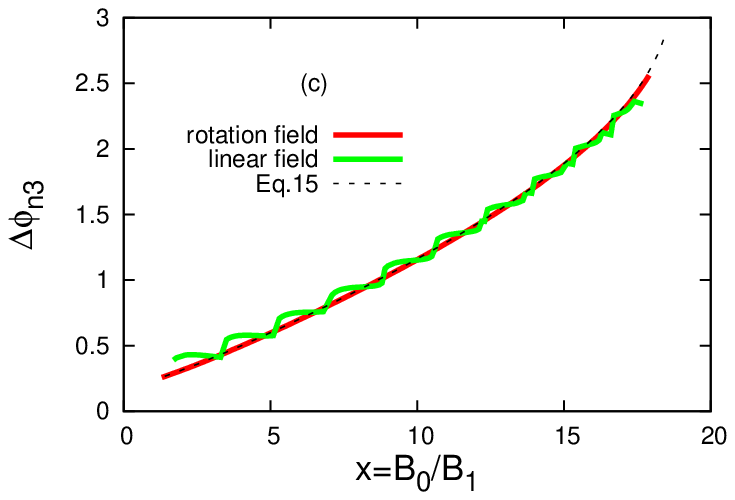}
 \caption{(Color figure) Solutions of the $\pi/2$ pulses for neutron and $^3$He, i.e., $R=\gamma_3 / \gamma_n$. The holding field $B_0$ is 10 mG along the $z$-axis. (a) $y=2\wrf /(\omega_{0,1}+\omega_{0,2})$ versus $x=B_0/B_1$. (b) Time duration $\tau$ versus $B_0/B_1$. (c) The angular difference, $\Delta \phi_{n3}$, between spins of neutron and $^3$He after the $\pi/2$ pulses. The black-dashed curves and the red-solid curves are solutions for rotating rf fields using the analytical equations (Eqs.~\ref{eq:tau1} and ~\ref{eq:phi}) and the Bloch equation, respectively. The results from these two approaches are identical. The green-dotted curves are obtained using Bloch equation simulation with linear rf fields.}
\label{fig:pi2.phi}
 \end{figure}

The solution for the specific case of rotating neutron and $^3$He simultaneously with a $\pi/2$ pulse, relevant for a proposed neutron EDM experiment~\cite{nEDM:2002}, is shown in Fig.~\ref{fig:pi2.phi}. The black-dashed curve in Fig.~\ref{fig:pi2.phi}(a) gives the solution for $y$ versus $x$ at $R=\gamma_3 / \gamma_n = 1.1121$, where $n$ and $3$ represent neutron and $^3$He, respectively. Figure~\ref{fig:pi2.phi}(b) and (c) also show solutions of the duration 
$\tau$ from Eq.~\ref{eq:tau1} and the angular difference of spins of two species after a $\pi/2$ rotation, $\Delta \phi_{n3}=\phi_n - \phi_3$, from Eq.~\ref{eq:phi}. Also shown in Fig.~\ref{fig:pi2.phi} are the solutions for linear rf fields, to be discussed next.

The study so far assumes a rotating rf field. However, a linear rf field is commonly utilized in experiments due to its simplicity to implement. Therefore, we extend our study to the case of a linear rf field, which can be decomposed in terms of two rotating components:
\begin{align}
B_{rf}(t)&=2B_1\cos(\wrf t)\uy \label{eq:brf}\\
&=B_1(\sin(\wrf t)\ux+\cos(\wrf t)\uy)+B_1(-\sin(\wrf t)\ux+\cos(\wrf t)\uy) 
\notag
\end{align}
The first component corresponds to the rotating field we have considered, 
which is stationary in the rotating frame. The second component 
rotates with a frequency of $-2\omega_{rf}$ in the rotating frame. 
To assess the effect of this high frequency term, we have solved the 
time dependence of $\vec S$ numerically using the Bloch 
equation (Eq.~\ref{eq:bloch}). We have considered the specific case for 
neutron and $^3$He with the gyromagnetic 
ratio $-18.32472$ and $-20.37895$ Hz/mG, respectively, and with $B_0$ at 10 mG. The Runge-Kutta method is applied and the time step of the simulation is $\Delta t = 10^{-6}$~sec. In order to determine a single $\pi/2$ pulse for both neutron and $^3$He, the following algorithm is  applied. Initially, both neutron and $^3$He spins are along the $\uz$-axis. Using the numerical simulation of the Bloch equation for a given $x=\frac{B_0}{B_1}$ and $y=\frac{2\omega_{rf}}{\omega_{0,1}+\omega_{0,2}}$, we derive the duration $\tau$ when the neutron spin is rotated into the $\ux-\uy$ plane, i.e., the $\uz$-component of the neutron spin becomes zero. In practical terms we derive the duration when the $\uz$-component of the neutron spin changes its sign, from a positive value to a negative value. Although the spin is wobbling, we just use the first point when the neutron spin is rotated into the $\ux-\uy$ plane. Then, at $\tau$, we consider the difference of the $\uz$-component between the spins of neutron and $^3$He. For a given $x=\frac{B_0}{B_1}$, we vary $y=\frac{2\omega_{rf}}{\omega_{0,1}+\omega_{0,2}}$ until the difference of the $\uz$-component between the spins of neutron and $^3$He changes sign. Using this algorithm, we can derive the solutions of the linear rf fields for the simultaneous $\pi/2$ rotation of neutron and $^3$He.

\begin{figure}[h]
\centering
\includegraphics[width=0.8\textwidth, height=0.45\textheight]{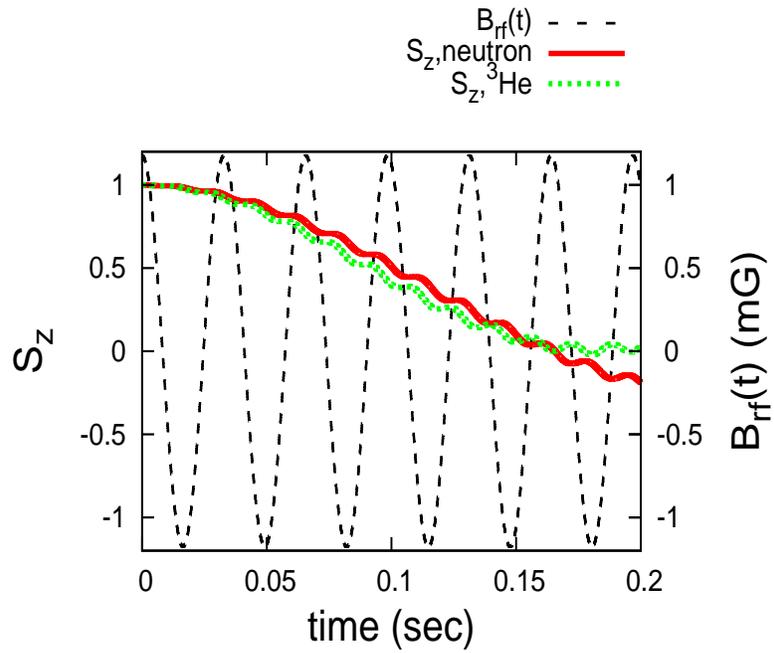}
\caption{(Color figure) The time dependence of $S_z$ for neutron and $^3$He with a linear rf field.  The red-solid and green-dotted curves are neutron and $^3$He, respectively. $B_0$ is 10 mG along the $\uz$-axis, and the black-dashed curve shows the amplitude and frequency of the linear rf field, $B_{rf}(t)$, defined in Eq.~\ref{eq:brf}. ($x=B_0/B_1=17$, $y=\frac{2\wrf}{\omega_{0,1}+\omega_{0,2}}=0.989596$)}
\label{fig:szn3}
\end{figure}

Fig.~\ref{fig:szn3} shows an example of a $\pi/2$ pulse for neutron and $^3$He using a linear rf field. The vertical components of the spins of the two species
as a function of time are shown as red-solid and green-dotted curves for neutron and $^3$He, respectively. The oscillatory pattern in these two curves has a frequency of $\sim 2\omega_{rf}$. Fig.~\ref{fig:szn3} shows that the both spins can be simultaneously rotated to the $\ux-\uy$ plane with the linear rf field.

In Fig.~\ref{fig:pi2.phi}(a), the solutions using the Bloch equation are shown for both rotating and linear rf fields. Significant difference between the two cases is observed for smaller values of $B_0/B_1$. At large values of $B_0/B_1$, the effect of the high-frequency counter rotating term could be neglected. The time duration $\tau$ and the angular difference after $\pi/2$ pulses for $R=\gamma_3/\gamma_n$ are also shown in Fig.~\ref{fig:pi2.phi}(b) and (c). They show that the results of the time duration in Eq.~\ref{eq:tau1} and the angular difference in Eq.~\ref{eq:phi} are consistent with those of numerical 
simulation using Bloch equation for rotating rf fields. The results for linear rf fields follow those for rotating rf fields, but with an oscillatory pattern superimposed. 

In summary, we have studied the solutions for $\pi/2$ pulses for two spin 
species of the same-sign gyromagnetic ratios using rotating and linear rf 
fields. The characteristics of the solutions are presented. For a specific 
experiment, the selection of the optimal values of $B_0$, $B_1$ could 
depend on various considerations. This study provides the solution 
for $\omega_{rf}$, once the values of $B_0$, $B_1$ are chosen. The 
application of this work to a neutron EDM experiment~\cite{nEDM:2002} is also 
discussed. Finally, this study can be extended to the case of two spin 
species having gyromagnetic ratios of opposite signs. 

We gratefully acknowledge valuable discussions with Bradley W. Filippone and Riccardo Schmid. This work was supported by the U.S. National Science Foundation and the Department of Energy.

\end{document}